\documentclass[aps, prl, superscriptaddress, preprintnumbers, nofootinbib, twocolumn]{revtex4}
\usepackage{amsmath}	
\usepackage{amsthm}	
\usepackage{amssymb}	
\usepackage{bbm} 
\usepackage{datetime}
\usepackage{graphicx}
\usepackage{color}
\usepackage{verbatim}

\usepackage{hyperref}

\definecolor{grey}{rgb}{0.4,0.4,0.4}
\definecolor{dullmagenta}{rgb}{0.4,0,0.4}
\definecolor{darkblue}{rgb}{0,0,0.4}
\definecolor{midblue}{rgb}{0,0,0.5}
\definecolor{midred}{rgb}{0.5,0,0}
\definecolor{orange}{rgb}{1,0.5,0}
\definecolor{lightbrown}{rgb}{0.75,0.5,0.25}
\definecolor{tan}{cmyk}{0.14,0.42,0.56,0}
\definecolor{djunglegreen}{cmyk}{0.99,0,0.52,0}
\definecolor{lightgreen}{rgb}{0,1,0}
\definecolor{olivegreen}{cmyk}{0.64,0,0.95,0.40}
\definecolor{midgreen}{rgb}{0.0,0.675,0.0}
\definecolor{darkgreen}{rgb}{0,0.5,0}

\newcommand{\q}{\quad}

\newcommand{\vs}{\vspace}

\renewcommand{\.}{\hspace{0.5mm}}

\newcommand{\ra}{\ensuremath{\rightarrow}}

\newcommand{\Ra}{\ensuremath{\Rightarrow}}

\newcommand{\Grm}{\ensuremath{\mathrm{G}}}

\newcommand{\Vrm}{\ensuremath{\mathrm{V}}}

\newcommand{\crm}{\ensuremath{\mathrm{c}}}

\newcommand{\hrm}{\ensuremath{\mathrm{h}}}

\newcommand{\jrm}{\ensuremath{\mathrm{j}}}

\newcommand{\Dcal}{\ensuremath{\mathcal{D}}}

\newcommand{\Hcal}{\ensuremath{\mathcal{H}}}

\newcommand{\Ocal}{\ensuremath{\mathcal{O}}}
\newcommand{\Pcal}{\ensuremath{\mathcal{P}}}

\newcommand{\Scal}{\ensuremath{\mathcal{S}}}

\newcommand{\Zcal}{\ensuremath{\mathcal{Z}}}

\newcommand{\onebbm}{\ensuremath{\mathbbm{1}}}

\newcommand{\kbm}{\ensuremath{\bm{k}}}

\renewcommand{\d}{\ensuremath{\mathrm{d}}}

\newcommand{\juu}{{1\hspace{-0.54ex}\mathrm{l}\hspace{-0.30ex}\mathrm{l}}} 

\newcommand{\Tr}{\ensuremath{\mathrm{Tr}}}

\newcommand{\eg}{e.g.}
\newcommand{\ie}{i.e.}

\newcommand{\cf}{cf.} 

\let\baraccent=\= 
\renewcommand{\=}[1]{\stackrel{#1}{=}} 

\theoremstyle{definition}

\theoremstyle{remark}

\smallskipamount

\settimeformat{ampmtime}

\usepackage{float}

\begin{document}

\title{On Stochastic Effects and Primordial Black-Hole Formation}

\author{Florian K{\"u}hnel}
\email{florian.kuhnel@fysik.su.se}
\affiliation{The Oskar Klein Centre for Cosmoparticle Physics,
	Department of Physics,
	Stockholm University,
	AlbaNova University Center,
	Roslagstullsbacken 21,
	SE--106\.91 Stockholm,
	Sweden}

\author{Katherine Freese}
\email{ktfreese@umich.edu}
\affiliation{Department of Physics,
	University of Michigan,
	Ann Arbor,
	MI 48109,
	USA}

\affiliation{The Oskar Klein Centre for Cosmoparticle Physics,
	Department of Physics,
	Stockholm University,
	AlbaNova University Center,
	Roslagstullsbacken 21,
	SE--106\.91 Stockholm,
	Sweden}

\date{\formatdate{\day}{\month}{\year}, \currenttime}


\begin{abstract}
The effect of large quantum fluctuations on primordial black-hole formation for inflationary models with a quasi-inflection point is investigated. By using techniques of stochastic inflation in combination with replica field theory and the Feynman-Jensen variational method, it is non-perturbatively demonstrated that the abundance of primordial black holes is amplified by several orders of magnitude as compared to the classical computation.
\end{abstract}

\maketitle


Cosmological inflation \cite{Guth:1980zm, Starobinsky:1980te, Albrecht:1982wi, Linde:1981mu} is a fundamental building block of our current understanding of the Universe. In addition to explaining the flatness and homogeneity of the Cosmos, inflation predicts the generation of perturbations from quantum fluctuations in the early Universe. The most common way for a realization of inflation is via a single scalar field, the so-called inflaton $\varphi$. Quantum fluctuations of the latter as well as of the metric seed today's structure in the Universe. The predictions of inflation are in remarkable agreement with measurements (\cf~Refs.~\cite{2018arXiv180706209P, Ade:2015tva, Ade:2015lrj}).

Under certain circumstances, the quantum fluctuations of the inflation field can be dominant over its classical evolution. There are two important cases in which this happens. One is typically at larger values of the inflaton potential $\Vrm( \varphi )$, yielding eternally expanding patches of the Universe \cite{Vilenkin:1983xq, Starobinsky:1986fx, Linde:1986fd} (for a review see Ref.~\cite{Guth:2007ng}). The other case occurs when the inflaton potential possesses a (quasi-)inflection point or one (or a multiple) plateau-like feature. This can be realized in numerous scenarios, such as double inflation \cite{Kannike:2017bxn}, radiative plateau inflation \cite{Ballesteros:2017fsr}, or in string theory (\eg~Refs.~\cite{Cicoli:2018asa, Ozsoy:2018flq}).

Classically, using the slow-roll conditions,
\begin{align}
	| \overset{..}{\varphi} |
		&\ll
					3\.H\.| \overset{.}{\varphi} |
					\; ,
	\qquad
	( \overset{.}{\varphi} )^{2}
		\ll
					2\.\Vrm( \varphi )
					\; ,
					\label{eq:slow-roll-conditions}
\end{align}
where an overdot represents a derivative with regard to cosmic time $t$, $H \equiv \overset{.}{a} / a$ is the Hubble parameter, and $a$ is the scale factor, the Universe inflates as
\begin{align}
	N
		&=
					\int\d t\;H
		=
					\int\d\varphi\;\frac{ H }{ \overset{.}{\varphi} }
	\q\Ra\q
	\delta \varphi_{\rm classical}
		=
					\frac{ \overset{.}{\varphi} }{ H }
					\; .
					\label{eq:delta-phi-classical}
\end{align}
with $N$ being the number of e-folds. On the other hand, the corresponding quantum fluctuations are
\begin{align}
	\delta \varphi_{\rm quantum}
		&=
					\frac{ H }{ 2 \pi }
					\; .
					\label{eq:delta-phi-quantum}
\end{align}
In turn, noting that the primordial metric perturbation
\begin{align}
	\zeta
		&=
					\frac{ H }{ \overset{.}{\varphi} }\.
					\delta \varphi
		=
					\frac{ \delta \varphi_{\rm quantum} }
					{ \delta \varphi_{\rm classical} }
					\label{eq:zeta}
\end{align}
implies that quantum effects can expected to be important, whenever $\zeta$ becomes of order one. This is often the case in primordial black hole (PBH) formation \cite{ZeldovichNovikov69, Hawking:1971ei, Hawking:1974sw, Carr:1975qj}, which will be discussed further below.

\begin{figure}[t]
	\vs{-4.5mm}
	\includegraphics[scale=1,angle=0]{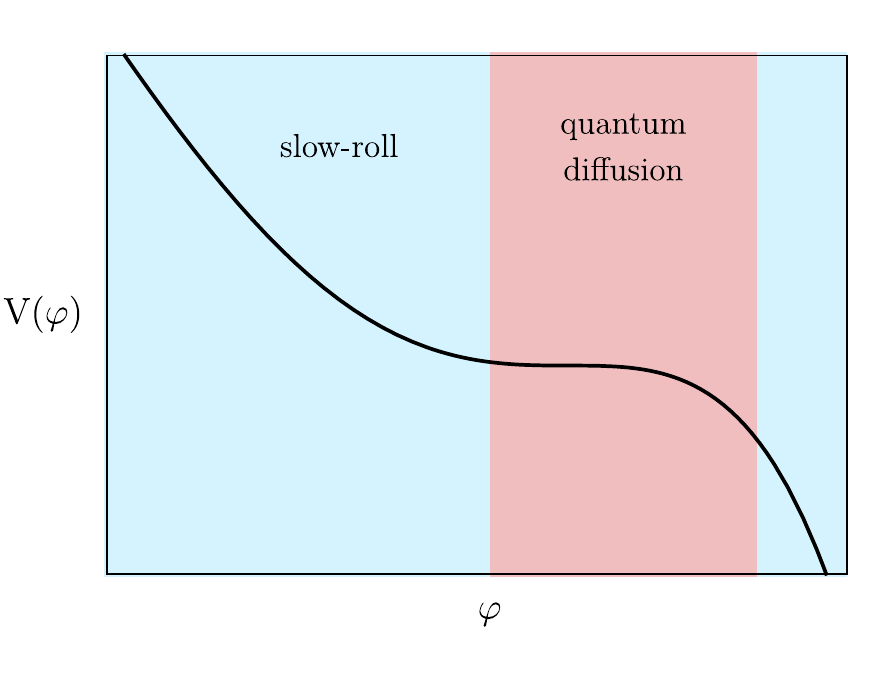}
	\vs{-10mm}	
	\caption{
		Sketch of an inflaton potential 
		with an inflection point.
		}
	\label{fig:V}
\end{figure}

Flatness of the inflaton potential, \ie~$\d \Vrm( \varphi ) / \d \varphi = 0$, leads to a growth of the primordial power spectrum of comoving curvature perturbations
\begin{align}
	\Pcal^{}_{\zeta}
		&=
					\frac{ H }{ 2 \pi\.\varphi' }
					\; ,
					\label{eq:P}
\end{align}
where $\varphi' \equiv \d \varphi / \d N$. The classical equation of motion,
\begin{align}
	\varphi''
	+
	3\.\varphi'
	+
	\frac{ 1 }{ H^{2} }
	\frac{\d \Vrm( \varphi )}{\d \varphi}
		&=
					0
					\; ,
					\label{eq:EOM-V'=0}
\end{align}
becomes\vs{-1mm}
\begin{align}
	\varphi''
	+
	3\.\varphi'
		=
					0
					\; ,
					\label{eq:EOM-V'=0}
\end{align}
which, by virtue of Eq.~\eqref{eq:P}, yields
\begin{align}
	\Pcal^{}_{\zeta}
		&\sim
					e^{3 N}
					\; .
					\label{eq:P-e-to-3N}
\end{align}

We note that, during the a plateau or inflection-point phase, the slow-roll conditions Eq.~\eqref{eq:slow-roll-conditions} are violated. Whilst these are tantamount to $| \varepsilon_{1, 2} | \ll 1$, with the slow-roll parameters $\varepsilon_{1} \equiv - \overset{.}{H} / H^{2}$ and $\varepsilon_{2} \equiv \overset{.}{\varepsilon}_{1} / ( \varepsilon_{1} H )$, here, the flatness of the potential implies $| \varepsilon_{2} | \simeq 6$ (\cf~Ref.~\cite{Dimopoulos:2017ged}). Furthermore, as will be discussed below, quantum effects will start to dominate.

In order to proceed and to demonstrate the strength of the quantum effects, we begin by choosing an inflaton potential with a quasi-inflection point as shown in Fig.~\ref{fig:V}. For definiteness, we will follow Ref.~\cite{Gong:2017qlj} and make the choice
\vs{-1mm}
\begin{align}
	V
	\left(
		\tilde{\varphi}
	\right)
		&=
					V_{0}\mspace{1mu}
					\begin{cases}
						1
						+
						\sum\limits_{i = 1}^{5}
						\lambda_{i}\.
						\tilde{\varphi}^{i}
							&
								\q
								\left(
									\tilde{\varphi} \ge 0
								\right)
								\\[4mm]
						1
						+
						\sum\limits_{i = 1}^{3}
						\lambda_{i}\.
						\tilde{\varphi}^{i}
							&
								\q
								\left(
									\tilde{\varphi} < 0
								\right)
					\end{cases}
					\; ,
					\label{eq:V}
\end{align}
where $\tilde{\varphi} \equiv \varphi / M_{\rm Pl}$, with $M_{\rm Pl}$ being the Planck mass, and we chose $\lambda_{1} = -\.0.0353553$, $\lambda_{2} = -\.0.0115783$, $\lambda_{3} = -\.0.00235702$, $\lambda_{4} = 728.239$, $\lambda_{5} = -\.11882.9$, as well as $V_{0} = 1.55 \times 10^{-10}$.

\begin{figure}
	\includegraphics[scale=0.92,angle=0]{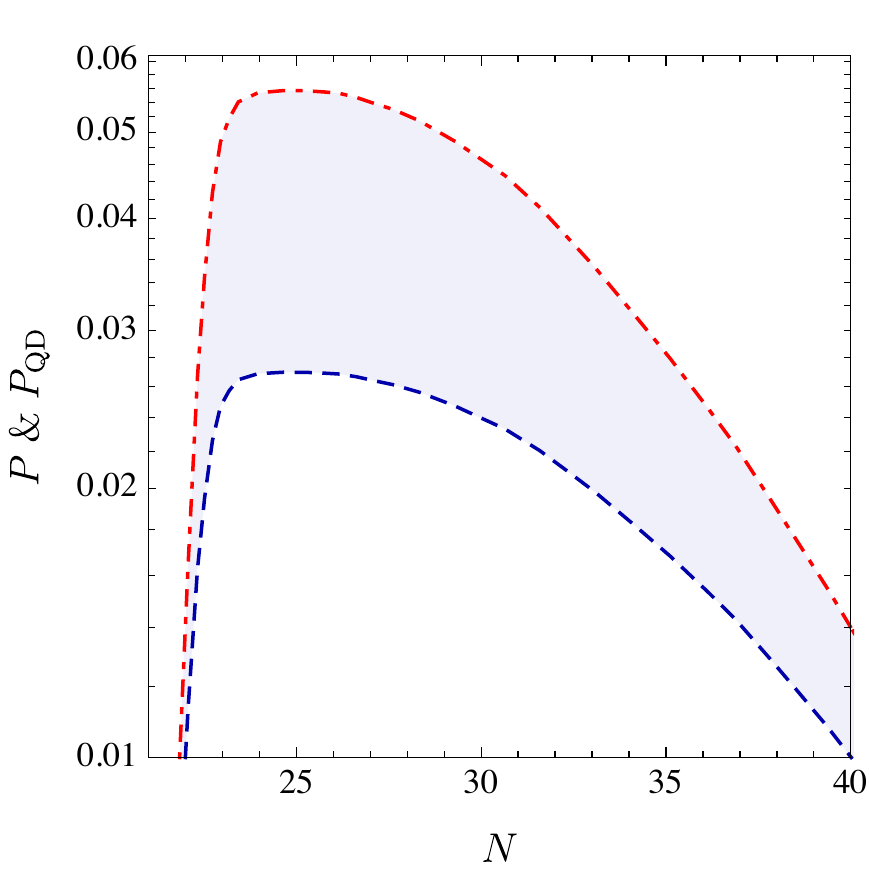}
	\caption{
		Dimensionless power spectra 
		$P$ (blue, dashed) and 
		$P_{\rm QD}$ (red, dot-dashed) 
		as a function of e-folds $N$.
		The upper curve includes stochastic effects.\\[3.02mm]
		}
	\label{fig:Power-Spectrum}
\end{figure}

In turn, the power spectrum can be calculated using the Mukhanov-Sasaki equation \cite{Mukhanov:1985rz, Sasaki:1986hm},
\begin{align}
	\frac{\d^{2} u_{k} }{\d\eta^{2}}
	+
	\left(
		k^{2}
		-
		\frac{ 1 }{ z }\.
		\frac{\d^{2} z }{ \d\eta^{2} }
	\right)
	u_{k}
		&=
					0
					\; ,
					\label{eq:Mukhanov-Sasaki-equation}
\end{align}
noting that $\Pcal_{\zeta} = ( 2\.\pi^{2} )^{-1}\.k^{3}\.| u_{k} / z |^{2}$. Above, $\eta$ denotes conformal time $\d\eta \equiv \d t / a$, $z \equiv a\.\Hcal^{-1}\.\d\phi / \d\eta = u_{k} / \zeta$, and $\Hcal = a^{-1}\.\d a / \d\eta$. The numerical (classical) solution to Eq.~\eqref{eq:Mukhanov-Sasaki-equation} is displayed in Fig.~\ref{fig:Power-Spectrum} (blue, dashed curved).

Of course, as mentioned above, quantum effects cannot be ignored during the inflection-point phase. In order to study these situations, Starobinsky introduced a stochastic framework \cite{Starobinsky:1986fx}. Its key idea lies in splitting the inflaton $\varphi$ into long- and short-wavelength modes, and viewing the former as classical objects evolving stochastically in an environment provided by quantum fluctuations of shorter wavelengths. Hence, it constitutes an example of how fundamental properties of quantum fields can be modelled using methods of statistical mechanics; one focusses on the ``relevant'' degrees of freedom (the long-wavelength modes) and regards the short-wavelength modes as ``irrelevant'' ones, where ``short'' and ``long'' are subject to the Hubble horizon.

In turn, the right-hand side of Eq.~\eqref{eq:Mukhanov-Sasaki-equation} acquires a stochastic source term $\hrm$, which is a Gaussian-distributed random variable with (see Ref.~\cite{Starobinsky:1986fx})
\begin{align}\label{eq:linear-distribution}
	\overline{\hrm}
		&=
					0
					\; ,
	\q
	\overline{\hrm^{2}}
		=
					\Delta
					\; ,
\end{align}
where $\Delta$ is a known function, depending on derivatives of the mode function (see \eg, Ref.~\cite{1987NuPhB.284..706R} for a detailed presentation). The overbar in Eq.~\eqref{eq:linear-distribution} and below denotes the average over the noise due to quantum fluctuations of short wavelengths.

Before we proceed, in order to provide an overview, let us briefly summarize the subsequent computational program (\cf~Ref.~\cite{Kuhnel:2010pp}): {\it 1.}~Wick-rotate to Euclidean signature, {\it 2.}~use the replica trick \cite{1988PhT....41l.109M}, in order to {\it 3.}~apply the Feynman-Jensen variational principle \cite{Feynman:1955zz}. This will allow us to go beyond ordinary perturbation theory (\cf~Ref.~\cite{1991JPhy1...1..809M}) and to obtain the full power spectrum, and therefore the PBH mass spectrum including stochastic modifications.

After Wick-rotating, we proceed with the replica trick (see the Appendix for a proof),
\begin{align}
	\overline{ \ln\big( \Zcal[ \mspace{1mu}\jrm\mspace{1mu} ] \big) }
		&=
					\lim_{m\mspace{1mu}\rightarrow\mspace{1mu}0}
					\frac{1}{m}
					\ln\!{
					\left(
						\overline{ \Zcal^{m}[ \mspace{1mu}\jrm\mspace{1mu} ] }
					\right)}
					\, ,
					\label{eq:relica-trick}
\end{align}
where $\Zcal[ \mspace{1mu}\jrm\mspace{1mu} ]$ is the generating functional depending on an external current $\jrm$. The different replica copies are labelled by the indices $a,\,b\mspace{1mu}$; their number is equal to $m$. We define the replicated action ${\Scal}^{(m)}$ via
\begin{align}
\begin{split}
	\overline{\Zcal^{m}[ \mspace{1mu}\jrm\mspace{1mu} ]}
		&=
					\int{\prod_{a\mspace{1mu}=\mspace{1mu}1}^{m}\!
					\Dcal[ \phi^{}_{a} ]}\,
					\overline{ \exp\!{ 
					\left(
						- \sum_{b\mspace{1mu}=\mspace{1mu}1}^{m}
						\Scal
						\big[
							\phi^{}_{b}, \jrm\mspace{1mu}
						\big]
					\right) } }\\[2.5mm]
		&\equiv
					\int{\prod_{a\mspace{1mu}=\mspace{1mu}1}^{m}\!
					\Dcal[ \phi^{}_{a} ]}\,
					\exp\!{
					\left(\!
						- {\Scal}^{(m)}
						\big[
							\{ \phi \}, \jrm\mspace{1mu}
						\big]
					\right) }
					\; .
					\label{eq:Zmbar}
\end{split}
\end{align}
Besides terms diagonal in replica space ($\propto\!\delta_{ab}$), it also contains the non-diagonal part
\begin{align}
	&{\Scal}^{(m)}\big[ \{ \phi \},\,\jrm\mspace{1mu} \big]
		\supset		
					-
					\frac{1}{2}\sum_{a,b\mspace{1mu}=\mspace{1mu}1}^{m}
					\int_{t, \kbm}\phi^{}_{a}( t, \kbm )\,
					{\Delta^{-1}}( t, \kbm )\,\phi^{}_{b}( t, -\kbm )
					\; ,
					\label{eq:S(m)}
					\\[-8mm]
					\notag
\end{align}
originating from the average over noise.

We apply the Feynman-Jensen variation principle, and therefore define a Gaussian variational action
\begin{equation}
	\Scal^{(m)}_{\rm var}\big[ \{ \phi \} \big]
		\equiv
					\frac{1}{2}\sum_{a,b\mspace{1mu}=\mspace{1mu}1}^{m}
					\int_{t, \kbm}\phi^{}_{a}( t, \kbm )\,
					{\Grm^{-1}}_{ab}( t, \kbm )\,\phi^{}_{b}( t, -\kbm )
					\; ,
\end{equation}
with $\int_{t} \equiv \int\d t$ and $\int_{\kbm} \equiv \int\d^{3} k / ( 2 \pi )^{3}$. The variational propagator $( \Grm_{ab} )_{a,b\mspace{1mu}=\mspace{1mu}1,\mspace{1mu}\ldots,\mspace{1mu}m}$ is a matrix in replica space, which is to be determined. In order to do so, we make the ansatz
\vs{-2mm}
\begin{align}
	{\Grm^{-1}}_{ab}
		&\equiv
					\Grm_{0}^{-1}\.\delta_{ab}
					-
					\sigma_{ab}
					\; .
					\label{eq:Gab-inverse-ansatz}
\end{align}
The self-energy matrix $\sigma$ mimics the diagonal and the non-diagonal parts in Eq.~\eqref{eq:Zmbar}.

Maximizing the right-hand side of the Feynman-Jensen inequality
\begin{align}
	\ln( \Zcal )
		&\ge
					\ln( \Zcal^{}_{\rm var} )
					+
					\Big\langle
						\Scal^{(m)}_{\rm var}
						-
						\Scal^{(m)}
					\Big\rangle_{\!\rm var}
					\; ,
					\label{eq:fvar}
\end{align}
wherein the subscript ``var'' refers to the variational action \eqref{eq:S(m)}, allows us to extract the replica structure $\sigma$, and in turn the full power spectrum. The latter is obtained via (\cf~Ref.~\cite{1991JPhy1...1..809M})
\begin{align}
	P_{\rm QD}( k )
		&=
					k^{3}
					\lim_{m\mspace{1mu}\ra\mspace{1mu}0}
					\frac{ 1 }{ m }\.
					\Tr
					\Big[\mspace{1mu}
						\Grm_{0}\.
						\onebbm
						+
						\Delta^{-1}\.
						\Grm_{0}^{2}\.\juu
					\mspace{1mu}\Big]
					\; ,
					\label{eq:Gphys}
\end{align}
where the matrices $\onebbm$ and $\juu$ are in replica space, with $\onebbm_{ab} = \delta_{ab}$, and $\juu_{ab} = 1$ for all $a,\,b$.

Figure \ref{fig:Power-Spectrum} shows the results for the full power spectrum $P_{\rm QD}( k )$ with quantum-diffusion effects (red, dot-dashed curve) in comparison to the power spectrum $P( k )$ without (blue, dashed curve). Both of these spectra are enhanced by the inflection point. As can be seen, the resummed stochastic effects yield an increase of power.

\begin{figure}
	\includegraphics[scale=0.92,angle=0]{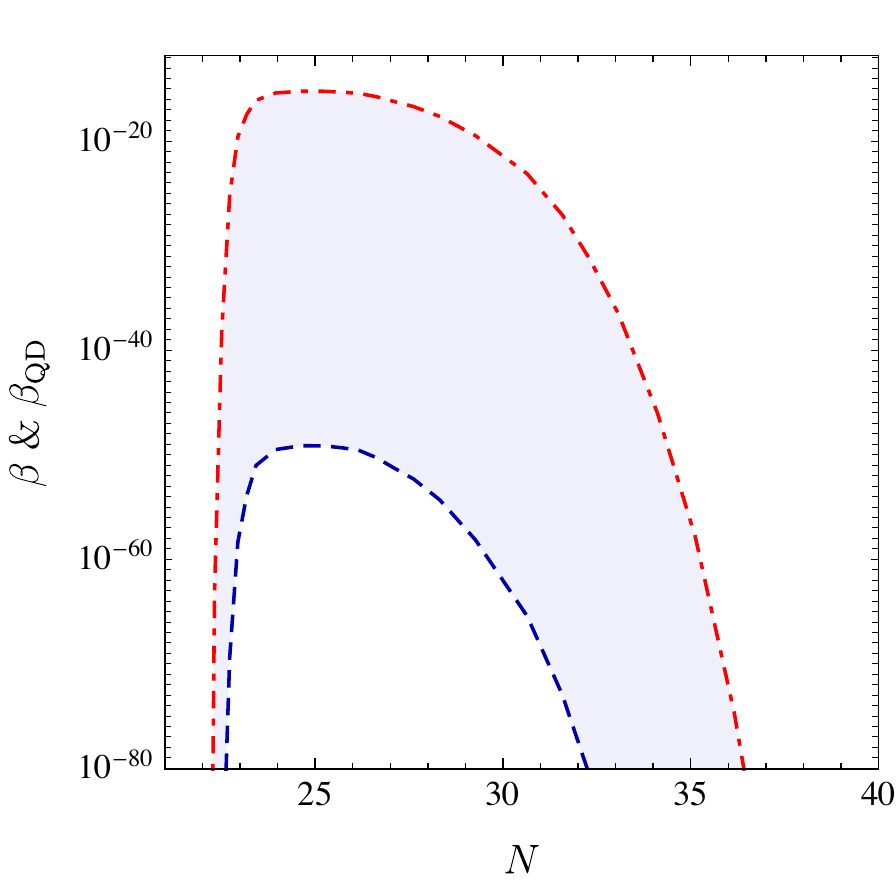}
	\caption{
		Fractions 
		$\beta$ (blue, dashed) and 
		$\beta_{\rm QD}$ (red, dot-dashed)
		of horizon patches collapsing to PBH
		as functions of e-folds $N$. 
		As in Fig.~\ref{fig:Power-Spectrum}, 
		the upper curve depicts the results of
		incorporating quantum diffusion.
		}
	\label{fig:beta}
\end{figure}

Having described a methodology to effectively incorporate the large quantum effects during the inflection-point phase, we are now in a position to derive the probability $\beta( M )$ that an overdense region of mass $M$ has a size exceeding the Jeans length at maximum expansion, so that it can collapse against the pressure and form a black hole. If the horizon-scale fluctuations have a Gaussian distribution with dispersion $\sigma$, \ie~root-mean square of the primordial power spectrum $P$, one expects for the fraction $\beta$ of horizon patches collapsing to a black hole \cite{Carr:1975qj}
\begin{align}
	&\beta
		\approx {\rm Erfc}\!
				\left[
					\frac{ \delta_{\crm} }
						{ \sqrt{2\.}\.\sigma }
				\right]
				.
				\label{eq:beta}
\end{align}
Here, Erfc is the complementary error function $\mathrm{Erfc} \equiv 1 - \mathrm{Erf}$, and $\delta_{\crm}$ is the density-contrast threshold for PBH formation. A simple analytic argument \cite{Carr:1975qj} suggests a value of $\delta_{\crm} \approx 1 / 3$, but more precise arguments{\;---\;}both numerical \cite{Musco:2012au} and analytical \cite{Harada:2013epa}{\;---\;}suggest a somewhat larger value: $\delta_{\crm} = 0.45$. We will use this value in remainder of our work. The result for $\beta$ are given in Fig.~\ref{fig:beta}, both with and without the effect of quantum diffusion. Due to the exponential dependence of $\beta$ on the power spectrum [see Eq.~\eqref{eq:beta}], already a moderat increase of the latter by less than an $\Ocal( 1 )$ factor, yields an increase of the former by more than 30 orders of magnitude!


Clearly, this has consequences for the formation of PBHs as possible dark-matter candidates (see Ref.~\cite{Carr:2016drx} for a review). In particular, the extendedness of the mass function generally allows for a larger PBH dark-matter fraction than monochromatic cases (see Refs.~\cite{Dolgov:1992pu, Yokoyama:1998xd, Musco:2012au, Kuhnel:2015vtw, Carr:2019kxo}). We should note that there is room for where the region of enhancement exactly lies. The closer towards the end of inflation it occurs, the lighter the holes are; but also, at the same time, the lower $\beta$ has to be. This is because during radiation domination its growth is approximately proportional to the scale factor. Hence, in model building, special care needs to be taken in order for quantum diffusion not to over-produce PBHs, and therefore rule out, the model.

In other scenarios and using different approaches, other recent works \cite{Pattison:2017mbe, Biagetti:2018pjj, Ezquiaga:2018gbw} also reach the conclusion that quantum diffusion leads to an increase of the power spectrum, and, in turn, of $\beta$. However, firstly, the scenario investigated in this work, can be viewed as a general prototype class of inflationary models in which quantum diffusion becomes relevant. Secondly, the utilized methods{\;---\;}the replica trick and the Feynman-Jensen variational principle{\;---\;}are non-perturbative in nature; the stochastic effects are effectively resummed into a generalised self-energy contribution.

We should note that these technique are not limited to linear noise potentials and could include arbitrary self-interactions (\cf~the original Ref.~\cite{1991JPhy1...1..809M} for the context of spin glasses, and Refs.~\cite{Kuhnel:2010pp, Kuhnel:2008wr, Kuhnel:2008yk, Kuhnel:2009zz} for its application to stochastic inflation). Furthermore, it can easily be applied to multi-field scenarios. Let us finally point out that the replica method used in this work has a close connection to the functional renormalisation group (\cf~Refs.~\cite{PhysRevB.52.1242, PhysRevB.68.174202}).
\vs{1mm}


\acknowledgments
The authors thank the anonymous referee for valuable remarks. They acknowledge support from DoE grant DE-SC0007859 at the University of Michigan, the Leinweber Center for Theoretical Physics, the Vetenskapsr{\aa}det (Swedish Research Council) through contract No.~638-2013-8993, and the Oskar Klein Centre for Cosmoparticle Physics.


\begin{appendix}
\section{Appendix}
Here, we give a proof of the replica trick\footnote{The replica trick is much older than one might think. According to Giogio Parisi (see Ref.~\cite{2005stgo.book.....G}) it can be dated back to at least the fourteenth century when the bishop of Lisieux applied a similar trick to define non-integer powers! The first non-trivial application of the replica trick to modern physics has been done by Edwards and Anderson in 1975 \cite{Edwards_1975}.}:
\begin{align}
\begin{split}
	\overline{ \ln( \Zcal ) }
		&=
					\lim_{m\mspace{1mu}\rightarrow\mspace{1mu}0}
					\frac{ m }{ m }\;
					\overline{ \ln( \Zcal ) }
					\\[1mm]
		&=
					\lim_{m\mspace{1mu}\rightarrow\mspace{1mu}0}
					\frac{1}{m}\.
					\ln\!{
					\left(
						1 + m\,\overline{ \ln( \Zcal ) }
					\right)}
					\\[1mm]
		&=
					\lim_{m\mspace{1mu}\rightarrow\mspace{1mu}0}
					\frac{1}{m}\.
					\ln\!{
					\left(
						\overline{ 1 + m \ln( \Zcal ) }
					\right)}
					\\[1mm]
		&=
					\lim_{m\mspace{1mu}\rightarrow\mspace{1mu}0}
					\frac{1}{m}\.
					\ln\!{
					\left(
						\overline{ \exp\!\big[ m \ln( \Zcal ) \big] }
					\right)}
					\\[1mm]
		&=
					\lim_{m\mspace{1mu}\rightarrow\mspace{1mu}0}
					\frac{1}{m}\.
					\ln\!{
					\left(
						\overline{ \exp\!\big[ \mspace{-1mu}\ln( \Zcal^{m} ) \big] }
					\right)}
					\\[1mm]
		&=
					\lim_{m\mspace{1mu}\rightarrow\mspace{1mu}0}
					\frac{1}{m}\,
					\ln\!{
					\Big(
						\overline{ \Zcal^{m} }
					\Big)}
					\label{eq:relica-trick-derivation}
\end{split}
					\\[-3mm]
		&
					\mspace{180mu}
					\Box
					\notag
\end{align}
\end{appendix}


\bibliography{refs}

\end{document}